\documentclass[superscriptaddress,twocolumn,showpacs,a4paper,
amssymb,amsmath,nobibnotes,aps,prd,longbibliography,
showkeys,nofootinbib,notitlepage]{revtex4-1}
\usepackage{verbatim}
\usepackage[T1]{fontenc}
\usepackage[utf8]{inputenc}
\usepackage[american]{babel}
\usepackage{epsfig}
\usepackage{graphicx,subfigure,bm,color,psfrag,hyperref}
\usepackage{booktabs}
\usepackage{multirow}
\usepackage{dcolumn}
\usepackage{amsmath}
\usepackage{mathtools}
\usepackage{amsfonts}
\usepackage{amssymb}
\usepackage{epstopdf}
\usepackage{bm}
\usepackage{siunitx}
\usepackage{braket}
\usepackage{enumitem}
\usepackage{soul}
\usepackage{todonotes}

\usepackage{color}

\makeatletter 
\providecommand*{\input@path}{}
\g@addto@macro\input@path{{"C:/Users/me/inpath/"}}
\makeatother

\definecolor{navyblue}{rgb}{0.0, 0.0, 0.5}
\definecolor{royalblue}{rgb}{0.25, 0.41, 0.88}
\definecolor{cadmiumgreen}{rgb}{0.0, 0.42, 0.24}
\definecolor{blue-violet}{rgb}{0.54, 0.17, 0.89}
\definecolor{darkviolet}{rgb}{0.58, 0.0, 0.83}
\definecolor{orange(colorwheel)}{rgb}{1.0, 0.5, 0.0}
\definecolor{greenW}{rgb}{0.0, 0.55, 0.1}

\usepackage{hyperref}
\hypersetup{
    colorlinks=true, 
    linkcolor=blue, 
    citecolor=magenta}

\usepackage{booktabs}
\usepackage{multirow}
\usepackage{dcolumn}
\usepackage{colortbl}

\begin{document}

\title{Coupled phantom cosmological model motivated by the warm inflationary paradigm}

\author{Sudip Halder}
\email{sudip.rs@presiuniv.ac.in}
\affiliation{Department of Mathematics, Presidency University, 86/1 College Street, Kolkata 700073, India}

\author{Supriya Pan}
\email{supriya.maths@presiuniv.ac.in}
\affiliation{Department of Mathematics, Presidency University, 86/1 College Street, Kolkata 700073, India}
\affiliation{Institute of Systems Science, Durban University of Technology, PO Box 1334, Durban 4000, Republic of South Africa}

\author{Paulo M.\ S\'a}
\email{pmsa@ualg.pt}
\affiliation{Departamento de F\'\i sica, Faculdade de Ci\^encias e Tecnologia, Universidade do Algarve, Campus de Gambelas, 8005-139 Faro, Portugal}
\affiliation{Instituto de Astrof\'\i sica e Ci\^encias do Espa\c co, Faculdade de Ci\^encias, Universidade de Lisboa, Campo Grande, 1749-016 Lisboa, Portugal}

\author{Tapan Saha}
\email{tapan.maths@presiuniv.ac.in}
\affiliation{Department of Mathematics, Presidency University, 86/1 College Street, Kolkata 700073, India}

\begin{abstract}
In this article, we investigate a coupled phantom dark-energy cosmological model in which the coupling term between a phantom scalar field with an exponential potential and a pressureless dark-matter fluid is motivated by the warm inflationary paradigm. Using methods of qualitative analysis of dynamical systems, complemented by numerical solutions of the evolution equations, we study the late-time behavior of our model. We show that contrary to the uncoupled scenario, the coupled phantom model admits accelerated scaling solutions. However, they do not correspond to a final state of the universe's evolution and, therefore, cannot be used to solve the cosmological coincidence problem. Furthermore, we show that, for certain coupling parameter values, the total  equation-of-state parameter's asymptotic behavior is significantly changed when compared to the uncoupled scenario, allowing for solutions less phantom even for steeper potentials of the phantom scalar field.
\end{abstract}
\maketitle

\section{Introduction \label{Sec-1}}

Modern cosmology has received tremendous attention from the scientific community due to the availability of a large number of astronomical probes.
The discovery of the cosmic microwave background radiation \cite{Penzias:1965wn} demanded a theory for our early universe, and inflation \cite{Guth:1980zm,Linde:1981mu} --- an accelerating expansion of the universe during its early time --- served as a potential proposal for explaining a number of early universe puzzles.
At the end of the 1990s, Supernovae Type Ia observations revealed that our universe is presently experiencing another phase of accelerating expansion \cite{SupernovaSearchTeam:1998fmf,SupernovaCosmologyProject:1998vns}.
This late accelerating expansion was further confirmed by other complementary observations \cite{Astier:2012ba}, and as a consequence, a theory for describing this phenomenon became essential. 

To explain the present-day accelerating expansion of the universe, two common approaches are usually put forward.
One is the introduction of some hypothetical dark energy (DE) fluid with large negative pressure in the context of Einstein's General Relativity (GR) \cite{Peebles:2002gy,Copeland:2006wr,Sahni:2006pa,Bamba:2012cp}.
Alternatively, modifying GR or introducing new gravitational theories beyond GR in various ways can explain this late-time accelerating expansion; such models are widely known as modified gravity (MG) models \cite{Nojiri:2006ri,Nojiri:2010wj,DeFelice:2010aj,Capozziello:2011et,Clifton:2011jh,Koyama:2015vza,Cai:2015emx,Nojiri:2017ncd,Bahamonde:2021gfp} and sometimes the resulting fluid in this sector mimicking the behavior of DE is known as geometrical DE.  
The concepts of DE and MG introduced plenty of cosmological models in the literature, which have been widely investigated with various astronomical probes \cite{Peebles:2002gy,Copeland:2006wr,Sahni:2006pa,Bamba:2012cp,Nojiri:2006ri,Nojiri:2010wj,DeFelice:2010aj,Capozziello:2011et,Clifton:2011jh,Koyama:2015vza,Cai:2015emx,Nojiri:2017ncd,Bahamonde:2021gfp}.
However, based on up-to-date observational evidences, the actual reason for this accelerating expansion --- DE, geometrical DE, or any other alternatives --- is not yet known. 
Additionally, a significant amount of non-luminous dark matter (DM), which is responsible for structure formation, exists in our universe.
A small amount of the total energy density ($\sim 4\%$) is contributed by baryons, photons, and neutrinos. 
Thus, the dynamics of our universe is dominated mainly by DM and DE (geometrical DE). 
Now, when considering a wide variety of cosmological scenarios accounting for both DM and DE (or geometrical DE), a large span of observational data is in favor of a simple cosmological scenario constructed within the context of GR plus a positive cosmological constant $\Lambda$, the so-called $\Lambda$CDM cosmological model.
In this model, DM is a pressureless nonrelativistic fluid (i.e., cold DM abbreviated as CDM) and $\Lambda$ serves as DE.
Additionally, in this cosmological setup, DE and DM each have their own conservation equations, meaning that they evolve independently with the expansion of the universe.
However, $\Lambda$CDM has faced some challenges in the past, such as the cosmological constant problem~\cite{Weinberg:1988cp} and the cosmic coincidence problem~\cite{Zlatev:1998tr}. 
Furthermore, according to recent observational data, cosmological tensions are also challenging the standard $\Lambda$CDM model, leading to the argument that this model is probably an approximate version of a more realistic theory which is not yet known \cite{Abdalla:2022yfr}.
Thus, an extension of the $\Lambda$CDM cosmology is welcome in order to tackle these problems. 

One of the generalizations of the $\Lambda$CDM cosmology is the theory of interacting DE or coupled DE where an interaction (i.e., energy exchange mechanism) between DM and DE is allowed.
Interacting cosmologies have many attractive consequences, e.g., the alleviation of the cosmic coincidence problem \cite{Amendola:1999er,Cai:2004dk,Pavon:2005yx,Huey:2004qv,delCampo:2008sr,delCampo:2008jx}, phantom crossing \cite{Wang:2005jx,Das:2005yj,Sadjadi:2006qb,Pan:2014afa}, and reconciling the cosmological tensions \cite{Kumar:2017dnp,DiValentino:2017iww,Yang:2018euj,Pan:2019gop,Pourtsidou:2016ico,An:2017crg,Kumar:2019wfs}.
The above interesting outcomes motivated many researchers to work on interacting cosmologies, and since the beginning of the 21st century to the present date, a multitude of interacting cosmological models have been studied~\cite{Amendola:1999er,Cai:2004dk,Yang:2018euj,Pan:2019gop,Barrow:2006hia,Valiviita:2008iv,Sa:2009,Gavela:2009cy,Valiviita:2009nu,Cao:2010fb,He:2011qn,Yang:2014gza,Li:2013bya,Yang:2014okp,Yang:2014hea,Pan:2012ki,Nunes:2016dlj,Yang:2016evp,Yang:2017yme,Yang:2017iew,Mifsud:2017fsy,Yang:2017ccc,Yang:2017zjs,Yang:2018uae,Pan:2020zza,Sa:2020a,Pan:2020mst,Sa:2020b,DiValentino:2019ffd,DiValentino:2020kpf,Sa:2021,Gao:2021xnk,Yang:2021hxg,Lucca:2021eqy,Sa:2022,Chatzidakis:2022mpf,Zhai:2023yny,Li:2023fdk,Teixeira:2023zjt,Sa:2023coi,Giare:2024ytc,Halder:2024uao,Giare:2024smz}.
The heart of interacting cosmologies is the coupling function or the interaction rate (also known as the interaction function) that controls the energy flow between the dark sectors.
As the interaction function modifies the evolution of the dark components at the background and perturbation levels, the choice of the interaction function is of great importance.

In the present article, we consider an interacting scenario between a phantom DE scalar field and a pressureless DM fluid in which the interaction is motivated by the warm inflationary paradigm\footnote{The cosmological model, in which a quintessence DE scalar field interacts directly with a pressureless DM fluid through a dissipative term inspired by warm inflation, was studied in Ref.~\cite{Sa:2023coi}.}.
We analyzed such a model using dynamical system techniques. Note that, because the total energy of the phantom field is unbounded from below, this model should be viewed as phenomenological, appropriate only to describe the late-time evolution of the universe. 

According to the warm-inflationary paradigm \cite{Berera:1995ie} (see also \cite{Berera:2023liv}), energy is continuously transferred from an inflaton field $\psi$ to a radiation bath, and hence, the energy density of the radiation sector, $\rho_\texttt{R}$, is not thinned out during the inflationary expansion.
As a result of this energy transfer, a post-inflationary radiation-dominated phase is found without the need for a reheating period, which is essential in the standard inflationary scenario~\cite{Guth:1980zm,Linde:1981mu}. 
Therefore, in the warm inflationary paradigm, assuming the well-known Friedmann-Lema\^{i}tre-Robertson-Walker (FLRW) geometry for the background, the evolution equations for the inflaton field and the radiation sector require a dissipative term as follows,
\begin{align}
 \dot{\psi} \ddot{\psi}+ 3H \dot{\psi}^2+  \frac{\partial V}{\partial\psi} \dot{\psi} &=-\Gamma\dot{\psi}^2,
\\ \dot{\rho}_\texttt{R}+4H\rho_\texttt{R} &= \Gamma \dot{\psi}^2,
\end{align}
where $H$ is the Hubble parameter, $V=V(\psi)$ denotes the potential of the inflaton field, and $\Gamma$ is the dissipation coefficient.
In general, $\Gamma$ might be a function of the inflaton field and the temperature $T$ of the radiation bath, meaning that $\Gamma=\Gamma(\psi,T)$. 
The warm-inflationary paradigm has received considerable attention from the scientific community with both positive and negative comments (see Ref.~\cite{Berera:2023liv} and the references therein).
Since most cosmological theories have been challenged, and this reveals indeed a fruitful progress of science, we avoid the criticisms on warm inflation and focus ourselves, in the present work, on the interacting dynamics in which the interaction function finds its motivation in the warm inflationary theory. 

This article is organized as follows.
In section~\ref{Sec-2}, we provide a detailed review of the uncoupled phantom DE cosmological model. Then, in section~\ref{Sec-3}, we present our coupled phantom DE cosmological model, in which the interaction term between DE and DM is inspired by the warm inflationary paradigm. For this model, we carry out a thorough dynamical system analysis and present the results. Finally, in section~\ref{Sec-conclusions}, we conclude the article by highlighting the key findings.

\section{Uncoupled phantom dark energy \label{Sec-2}}

In this section, the uncoupled phantom DE cosmological model is briefly reviewed (for more details, see \cite{Caldwell:1999ew,Schulz:2001yx,Gibbons:2003yj,Li:2003ft,Hao:2003th,Chimento:2003qy,Vikman:2004dc,Ludwick:2017tox} and the references therein).

We assume the flat FLRW metric that  takes the form 
\begin{eqnarray}\label{flrw}
 ds^2 = -dt^2 + a^2(t) d \Sigma^2 , 
\end{eqnarray}
where $a(t)$ denotes the expansion scale factor of the universe and $d \Sigma^2$ is the metric of the three-dimensional Euclidean space.

We further assume that the gravitational sector of the universe is described by Einstein's General Relativity (GR) and the matter sector, minimally coupled to gravity, comprises a pressureless DM fluid with energy density $\rho_{\rm DM}$ and a phantom DE scalar field $\phi$ with an exponential potential
\begin{equation}
V(\phi) = V_0 e^{-\lambda \kappa \phi},
 \label{potential}
\end{equation}
where $V_0$ and $\lambda$ are positive constants of dimension (mass)$^4$ and (mass)$^0$, respectively, and the notation $\kappa\equiv\sqrt{8\pi G} =\sqrt{8\pi}/m_p$
(here $m_p$ stands for the Planck mass) has been used.
We neglect radiation and baryons and their influence on the universe's late-time evolution.

Under the above assumptions, the evolution equations for the uncoupled phantom DE cosmological model become
\begin{gather}
H^2 = \frac{\kappa^2}{3} \left( - \frac{\dot{\phi}^2}{2} + V(\phi) + \rho_{\rm DM} \right),   \label{pde-1}
\\
\dot{H} = - \frac{\kappa^2}{2} \left(- \dot{\phi}^2 + \rho_{\rm DM}\right), \label{pde-2}
\\
\ddot{\phi} + 3H\dot{\phi}  - \frac{\partial V (\phi)}{\partial\phi} = 0,   \label{pde-3}
\\
\dot{\rho}_{\rm DM}+3H\rho_{\rm DM} = 0, 
\label{pde-4}
\end{gather}
where $H=\dot{a}/a$ is the Hubble parameter, an overdot denotes a derivative with respect to cosmic time $t$, and the energy density and pressure of the phantom scalar field are given by 
\begin{equation}
\rho_{\phi} = - \frac{\dot{\phi}^2}{2} + V (\phi) \quad
\mbox{and} \quad p_{\phi} = - \frac{\dot{\phi}^2}{2} - V (\phi),
	\label{rho_p}
\end{equation}  
respectively.  

Introducing the dimensionless variables
\begin{equation}
	x=\frac{\kappa \dot{\phi}}{\sqrt6 H} \quad
 \mbox{and} \quad
	y=\frac{\kappa\sqrt{V}}{\sqrt3 H},
	\label{variables pde}
\end{equation}
and a new time variable $\eta$, defined as
\begin{equation}
	\frac{d\eta}{dt}=H,	\label{eta}
\end{equation}
the above evolution equations yield the two-dimensional autonomous dynamical system
\begin{subequations} \label{DS-0}
\begin{align}
    x_\eta=&-\frac{\sqrt{6}}{2}\lambda y^2-\frac{3}{2} x 
    \left(1+x^2+y^2\right),   \label{DS-0-x} \\
    y_\eta=& \left[-\frac{\sqrt{6}}{2}\lambda x +\frac{3}{2}\left(1-x^2-y^2\right) \right] y, \label{DS-0-y} 
\end{align}
\end{subequations}
where the subscript $\eta$ denotes the derivative with respect to $\eta =\ln(a/a_0)$ {\color{red}and} $a_0$ refers to the present value of the scale factor. 
Note that the variable $\eta$ is nothing more than the number of $e$-folds $N$, a convenient measure of the expansion of the universe.

From the Friedmann equation (\ref{pde-1}), the DM density parameter $\Omega_{\rm DM}$, defined as the ratio between $\rho_{\rm DM}$ and the critical density $3H^2/\kappa^2$, can be expressed in terms of the dimensionless variables $x$ and $y$ as
\begin{equation}
\Omega_{\rm DM}=1+x^2-y^2,
	\label{Omega-matter}
\end{equation} 
and hence, the DE density parameter, defined as $\Omega_{\phi} = \kappa^2 \rho_{\phi}/3H^2$, can be expressed as 
\begin{equation}
    \Omega_\phi=-x^2+y^2= 1 - \Omega_{\rm DM}.
    \label{Omega-phi}
\end{equation}

Taking into account that the energy density and the pressure of the phantom scalar field are given by Eq.~(\ref{rho_p}), the phantom equation-of-state parameter $w_{\phi} = p_{\phi}/\rho_{\phi}$ and the total equation-of-state parameter $w_{\rm tot} = (p_{\phi}+p_{\rm DM})/(\rho_{\phi} + \rho_{\rm DM})$ can be expressed in terms of the dimensionless variables $x$ and $y$ as 
\begin{align}
w_{\phi} & =  \frac{x^2 +y^2}{x^2 - y^2},
    \label{w-phi}\\
w_{\rm tot} & = - (x^2 + y^2). 
    \label{w-tot}
\end{align}

Using the dynamical system (\ref{DS-0}), the evolution equation for the DM density parameter can be written as
\begin{equation}
    \Omega_{{\rm DM},\eta} = -3 \left(x^2+y^2\right) \Omega_{\rm DM}, \label{dm-evo-1}
\end{equation}
implying that the hyperbolas  $y=\pm\sqrt{1+x^2}$ are invariant manifolds, i.e., they are not crossed by phase-space orbits.
Inspection of Eq.~(\ref{DS-0-y}) further reveals that $y=0$ is also an invariant manifold. Since $\Omega_{\rm DM}$ is non-negative by definition ($y^2\leq 1+x^2$) and we are interested in non-contracting cosmological solutions ($y\geq0$), the phase space of the dynamical system (\ref{DS-0}) is given by
\begin{equation}
R_2=\left\{ (x,y)\in \mathbb{R}^2: y\leq\sqrt{1+x^2} , y \geq 0 \right\}.
    \label{phase space uncoupled}
\end{equation}

Note that the phantom equation-of-state parameter (\ref{w-phi}) becomes infinite for $y^2=x^2$. 
This is a direct consequence of the fact that, due to the negative sign in the kinetic-energy term, the total energy of the scalar field is no longer bounded from below, implying, from a quantum point of view, the appearance of ghosts in the theory, and, from a classical perspective, the instability of the equation-of-motion solutions under small perturbations~\cite{Carroll:2003st}.
After analyzing the stability of the critical points of the dynamical system (\ref{DS-0}) and describing the phase-space orbits, we shall return to this issue.

The dynamical system (\ref{DS-0}), in the finite region of the phase space, has just two critical points, $A_0$ and $A_1$.
Their properties (existence, eigenvalues, stability, and various cosmological features) are highlighted in Table \ref{first-table-unc-potI}.


\begin{table*}[]
\centering
	\begin{tabular}{c c c c c c c c}\hline\hline
Critical point & Existence  & Eigenvalues & Stability & $\Omega_\phi$ & $\Omega_{\rm DM}$ & $w_{\rm tot}$ & Acceleration\\ \hline
$A_0(0,0)$ & Always  & $\left(-\frac{3}{2},\frac{3}{2}\right)$   &  Saddle  &  0  &  1  &  0 & Never     \\
$A_1\left(-\frac{\lambda}{\sqrt{6}},\sqrt{1+\frac{\lambda^2}{6}}\right)$ & Always  & $\left(-3-\lambda^2,-3-\frac{\lambda^2}{2}\right)$   &  Attractor  &  1  &  0  &  $-1-\frac{\lambda^2}{3}$ & Always    \\
		 \hline\hline
	\end{tabular}%
	\caption{Critical points of the dynamical system (\ref{DS-0}) and their properties for the uncoupled phantom DE cosmological model.   }
	\label{first-table-unc-potI}
\end{table*}

The critical point $A_0$ is located at the origin $(0,0)$ of the phase space and always exists independently of the $\lambda$ parameter's value.
It represents a matter-dominated cosmological solution ($\Omega_{\rm DM}=1$) with decelerated expansion ($w_{\rm tot} =0$).
Since the eigenvalues of the Jacobian matrix of the dynamical system~(\ref{DS-0}) are nonzero and have opposite signs, linear stability theory indicates that $A_0$ is a saddle point.
In summary, this critical point corresponds to a DM-dominated decelerated phase of the universe's evolution.

The critical point $A_1$ also exists for any value of the parameter $\lambda$.
It lies on the hyperbola  $y=\sqrt{1+x^2}$ and represents a solution completely dominated by DE ($\Omega_\phi=1$).
Because $w_{\rm tot}<-1/3$ for any value of $\lambda$, it always corresponds to an accelerating solution.
Since both the eigenvalues are negative, $A_1$ is a global attractor.
Therefore, this critical point corresponds to a DE-dominated late-time accelerating solution. 

\begin{figure}
	\includegraphics[width=0.48\textwidth]{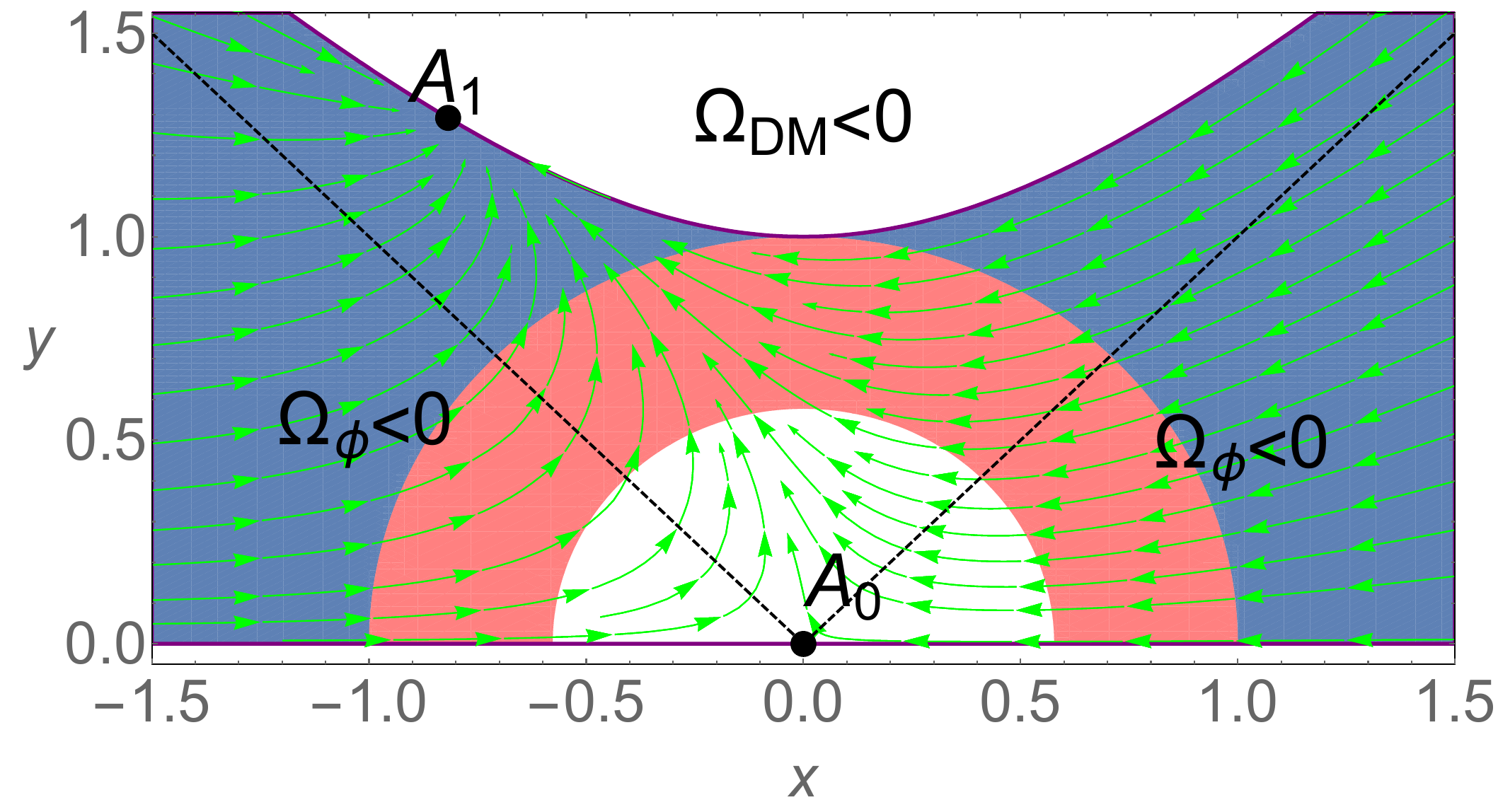}
	\caption{
	\label{fig1:uncoupled phantom}
Phase portrait of the dynamical system (\ref{DS-0}). $A_0$ and $A_1$ denote the DM-dominated and DE-dominated critical points, respectively. In color/shaded highlighted regions, the universe's expansion is accelerated, with $w_{tot}<-1$ (blue color, dark shading) and $-1<w_{tot}<-1/3$ (red color, light shading). Orbits starting near $A_0$ and converging to $A_1$  correspond to the universe's transition from a past decelerating matter-dominated phase to the present accelerating phase. The dotted black lines correspond to $y = \pm x$ where the phantom equation-of-state parameter $w_{\phi}$ diverges. In this figure, we have considered $\lambda=2$; for other values of this parameter, one would get similar graphics.
}
\end{figure}

To reproduce the succession of cosmological eras observed in the late-time evolution of the universe, namely, a matter-dominated era long enough to allow for structure formation followed by an accelerated era dominated by phantom  DE, phase-space orbits must first approach the critical point $A_0$, staying close to it long enough, and only then head to the critical point $A_1$ (see Fig.~\ref{fig1:uncoupled phantom}).

Note, however, that such orbits, coming from an infinitely far region of the phase space, obligatory cross the lines $x=y$ or $x=-y$, where the phantom equation-of-state parameter (\ref{w-phi}) diverges. 
Therefore, we shall attribute physical meaning to the solution only after the occurrence of this singularity, implying that the phantom DE model should be viewed as a phenomenological model describing the late-time evolution of the universe \cite{Bahamonde:2017ize}. 

To conclude this section, let us point out that the phase space of the dynamical system~(\ref{DS-0}) can be compactified (see, for instance, Ref.~\cite{Bahamonde:2017ize}).
However, because of the circumstances described in the previous paragraph, such a procedure is not needed to fully understand the behavior of the orbits of cosmological relevance.

\section{Coupled phantom dark energy motivated by warm inflation \label{Sec-3}}

Let us now turn to the coupled phantom DE cosmological model.
The cosmological scenarios in which a phantom DE scalar field directly interacts with the DM component generalize the uncoupled phantom DE  model presented in section~\ref{Sec-2}.
The interaction function characterizing the energy transfer between the phantom DE and DM plays the key role in this context.
Given that there is currently no fundamental theory that specifies the exact form of the interaction function between DE and DM, one must resort to a phenomenological approach, considering different couplings with different physical motivations.
In this work, we shall consider an interaction function motivated by the warm inflationary scenario.
Other choices of the interaction function have been considered in Refs.~\cite{Fu:2008gh,Chen:2008ft}.

For the coupled phantom  DE cosmological model, the evolution equations are the same as for the uncoupled case of section~\ref{Sec-2}, with the exception that Eqs.~(\ref{pde-3}) and (\ref{pde-4}) now become
\begin{align}
\ddot{\phi} + 3H\dot{\phi}  - \frac{\partial V}{\partial\phi}
&= - \frac{Q}{\dot{\phi}},   \label{cons-de}\\
\dot{\rho}_{\rm DM}+3H\rho_{\rm DM} &= -Q.  \label{cons-dm}
\end{align}
where $Q$ is the interaction term between the phantom DE scalar field and the DM fluid, determining the energy flow between them.
For $Q > 0$, the energy flows from  DM to phantom DE, while $Q < 0$ indicates an energy flow in the opposite direction, i.e., from the phantom scalar field to the DM fluid.

As in section~\ref{Sec-2}, we assume the phantom scalar field to have the exponential potential given by Eq.~(\ref{potential}).

Inspired by warm inflation, we choose the coupling between DE and DM to be of the form \cite{Sa:2023coi}
\begin{equation}
	Q=\Gamma \dot{\phi}^2,
	\label{interaction}
\end{equation}
where $\Gamma$ is a nonzero constant having the dimension of the Hubble rate.   

Let us now write the evolution equations, Eqs.~(\ref{pde-1}),  (\ref{pde-2}), (\ref{cons-de}), and (\ref{cons-dm}) for the coupled phantom  DE cosmological model as a dynamical system.
Since the interaction term $Q$ cannot be written as a function of the dimensionless variables of $x$ and $y$, introduced in Eq.~(\ref{variables pde}), one extra variable $z$ is needed to close the dynamical system, which, therefore, becomes three dimensional.
We choose  this extra variable to be \cite{Sa:2023coi} 
\begin{equation}
	z=\frac{H_*}{H \Omega_{\rm DM}+H_*},
	\label{variables}
\end{equation}
where $\Omega_{\rm DM}$ is the DM density parameter and $H_*$ is a positive constant representing the Hubble parameter at some particular instant $t = t_{*}$.

This choice of $z$ compactifies the phase space in the $z$ direction, between $z=0$ (for $H=+\infty$) and $z=1$ (for $H=0$), but it also introduces a singular term on the evolution equation for $x$, namely, the interaction term becomes proportional to $(1-z)^{-1}$, which diverges for $z\rightarrow1$.
To remove this singularity, we choose a new time variable $\tau$ as \cite{Sa:2023coi}
\begin{equation}
	\frac{d\tau}{dt}=\frac{H}{1-z}.
	\label{tau}
\end{equation}
Note that, due to the factor $1-z$, the variable $\tau$ does not have a simple physical interpretation as in the uncoupled case, where, recall, the variable $\eta$, given by Eq.~(\ref{eta}) was simply the $e$-fold number.

In the variables $x$, $y$, $z$, and $\tau$, the evolution equations~(\ref{pde-1}), (\ref{pde-2}), (\ref{cons-de}), and (\ref{cons-dm}) for the coupled phantom DE cosmological model give rise to the three-dimensional dynamical system  
\begin{subequations} \label{DS-1}
\begin{align}
x_\tau &= \left[ - \frac{\sqrt{6}}{2} \lambda y^2 
	- \frac{3}{2} x (1+x^2+y^2) \right] (1-z) \nonumber 
\\ 
		& \hspace{3mm} - \alpha x (1+x^2 - y^2) z, \label{aut-sys-potI-1}
\\
y_\tau &= \left[- \frac{\sqrt{6}}{2} \lambda x + \frac{3}{2}
(1-x^2 -y^2) \right] y (1-z), \label{aut-sys-potI-2}
\\
z_\tau &= \left[\frac{3}{2} (1 + x^2 +y^2) (1-z) + 2 \alpha x^2 z \right]
z (1-z), \label{aut-sys-potI-3}
\end{align}
\end{subequations}
where the subscript $\tau$ denotes a derivative with respect to this variable and $\alpha = \Gamma/H_{*}$ is the dimensionless coupling parameter, taken to be nonzero.
Following the convention of the direction of the energy transfer between the dark sectors as described above, $\alpha >0$ indicates the energy transfer from the DM sector to the phantom scalar field and $\alpha <0$ corresponds to the energy transfer in the reverse direction.

Note that the above dynamical system is invariant under the transformation $x\rightarrow-x$ and $\lambda \rightarrow -\lambda$, allowing us to assume, without any loss of generality, that the parameter $\lambda$ is positive.
Therefore, the parameter space of our coupled phantom DE model is $\{ (\alpha,\lambda): \alpha\neq0,\lambda>0\}$.

In what concerns the DM density parameter $\Omega_{\rm DM}$, the DE density parameter $\Omega_{\phi}$, the phantom equation-of-state parameter $w_{\phi}$, and the total equation-of-state parameter $w_{\rm tot}$, they do not depend on the variable $z$ and, therefore, are given by the same expressions as in the uncoupled case, namely, by Eqs.~(\ref{Omega-matter}), (\ref{Omega-phi}), (\ref{w-phi}), and (\ref{w-tot}), respectively.  

Inspection of the dynamical system~(\ref{DS-1}) and the evolution equation for the DM density parameter, 
\begin{equation}
    \Omega_{{\rm DM},\tau}=\Omega_{\rm DM} \left[ -3(x^2+y^2)(1-z)-2\alpha x^2 z \right],
\end{equation}
shows that the surfaces $y=0$, $z=0$, $z=1$, and $y=\pm\sqrt{1+x^2}$  are invariant manifolds.
Since we are not interested in contracting cosmologies (for which $H<0$, implying $y<0$) and taking into account that $\Omega_{\rm DM}\geq0$, the phase space of the dynamical system should be restricted to the region
\begin{equation}
R_3=\left\{ (x,y,z) \in \mathbb{R}^3: y\leq\sqrt{1+x^2}, y\geq0, 0\leq z\leq1 \right\}. \label{R3}
\end{equation}
This region, however, contains the surfaces $x = \pm y$, at which $w_\phi$ diverges, and, furthermore, for $y^2<x^2$ the DE density parameter $\Omega_\phi$ becomes negative.
To avoid these unphysical situations, we should further restrict the phase space to the region
\begin{eqnarray}
\overline{R}_3 =\Bigl\{ (x,y,z) \in \mathbb{R}^3: y\leq\sqrt{1+x^2}, y^2\geq x^2,\nonumber\\ y \geq 0, 0\leq z\leq1 \Bigr\},
\end{eqnarray}
meaning that our model is only suitable to describe the late-time evolution of the universe, that is, the part of the evolution occurring after the orbits cross the surfaces $x=y$ or $x=-y$.
This is the same restriction we have already encountered in the uncoupled scenario (see discussion in section~\ref{Sec-2}).   

In the finite region of the phase space\footnote{As in the uncoupled case, we do not need to study the dynamical system's behavior at infinity to describe the solutions of cosmological relevance. See discussion at the end of section~\ref{Sec-2}.}, the dynamical system (\ref{DS-1}) has three critical points, $A$, $B$, and $C$, and two critical lines, $D$ and $E$.
The qualitative features and the eigenvalues of these critical points are displayed in Table~\ref{first-table-potI} and Table~\ref{second-table-potI}, respectively.

\begin{table*}[t]
\centering
	\begin{tabular}{c c c c c c }\hline\hline
Critical point/line & Existence & $\Omega_\phi$ & $\Omega_{\rm DM}$ & $w_{\rm tot}$ & Acceleration \\ \hline
$A(0,0,0)$  & Always    &  0  &  1  &  0  &  Never \\
$B\left(-\frac{\lambda}{\sqrt{6}},\sqrt{1+\frac{\lambda^2}{6}},0\right)$  & Always   & 1   &  0  &  $-1-\frac{\lambda^2}{3}$  &  Always \\
$C\left(-\frac{\lambda}{\sqrt{6}},\sqrt{1+\frac{\lambda^2}{6}},\frac{3(6+\lambda^2)}{3(6+\lambda^2)-2\alpha\lambda^2}\right)$  & $\alpha<0$   & 1   &  0  & $-1-\frac{\lambda^2}{3}$   &  Always \\
$D(0,y,1)$  &  $0\leq y \leq 1$  &  $y^2$  & $1-y^2$   &  $-y^2$  &  $\frac{1}{\sqrt{3}}<y\leq 1$ \\
$E\left(x,\sqrt{1+x^2},1\right)$  &  Always  &  1  &  0  &  $-1-2x^2$  &  Always     \\
\hline\hline
\end{tabular}%
\caption{Properties of the critical points and critical lines of the dynamical system (\ref{DS-1}) for the coupled phantom DE cosmological model.
	   }
	\label{first-table-potI}
\end{table*}


\begin{table*}[t]
\centering
	\begin{tabular}{c c c}\hline\hline
Critical point/line & Eigenvalues & Stability\\ \hline
$A(0,0,0)$ & $\left(-\frac{3}{2},\frac{3}{2},\frac{3}{2}\right)$ & Saddle ($\alpha\neq0$) \\
$B\left(-\frac{\lambda}{\sqrt{6}},\sqrt{1+\frac{\lambda^2}{6}},0\right)$  & $\left(-3-\lambda^2,-3-\frac{\lambda^2}{2},3+\frac{\lambda^2}{2}\right)$ & Saddle ($\alpha\neq0$)         \\
$C\left(-\frac{\lambda}{\sqrt{6}},\sqrt{1+\frac{\lambda^2}{6}},\frac{3(6+\lambda^2)}{3(6+\lambda^2)-2\alpha\lambda^2}\right)$  & $\left(\frac{\alpha\lambda^4}{3\left(6+\lambda^2\right)-2\alpha\lambda^2},\frac{\alpha\lambda^2\left(6+\lambda^2\right)}{3\left(6+\lambda^2\right)-2\alpha\lambda^2},\frac{\alpha\lambda^2\left(6+\lambda^2\right)}{3\left(6+\lambda^2\right)-2\alpha\lambda^2}\right)$   & Attractor $(\alpha<0)$       \\
$D(0,y,1)$  & $\left(0,0,-\alpha\left(1-y^2\right)\right)$ &  Attractor ($\alpha>0$, $y=1$), Saddle (otherwise) \\
$E\left(x,\sqrt{1+x^2},1\right)$  & $\left(0,-2\alpha x^2,-2\alpha x^2\right)$ &  Attractor ($\alpha>0$, $\forall x$)   \\
\hline\hline
\end{tabular}%
	\caption{Stability of the critical points and critical lines of the dynamical system (\ref{DS-1}) for the coupled phantom DE cosmological model.}
	\label{second-table-potI}
\end{table*}

The critical points $A$ and $B$ were already present in the uncoupled case. However, point $B$ is not anymore an attractor, but rather a saddle point. The critical point $C$, as well as the critical lines $D$ and $E$, are new, arising due to the introduction of a direct coupling between the phantom DE scalar field and the DM fluid.

The critical point $A$ always exists for all allowed values of the model parameters $\alpha$ and $\lambda$.
It corresponds to a matter-dominated decelerating solution ($\Omega_{\rm DM}=1$ and $w_{\rm tot}=0$).
Since one eigenvalue of the Jacobian matrix of the dynamical system~(\ref{DS-1}) is negative and the other two are positive, linear stability theory indicates that $A$ is unstable, more specifically, it is a saddle point.

The critical point $B$ is also always present, representing a DE-dominated accelerating solution ($\Omega_\phi=1$, $w_{\rm tot} < -1$).
Because two eigenvalues are negative and one positive, $B$ is also a saddle point: all orbits approaching it near the $\{x,y\}$ plane are repelled along the $z$ direction.

\begin{figure}
    \centering
    \includegraphics[width=0.5\textwidth]{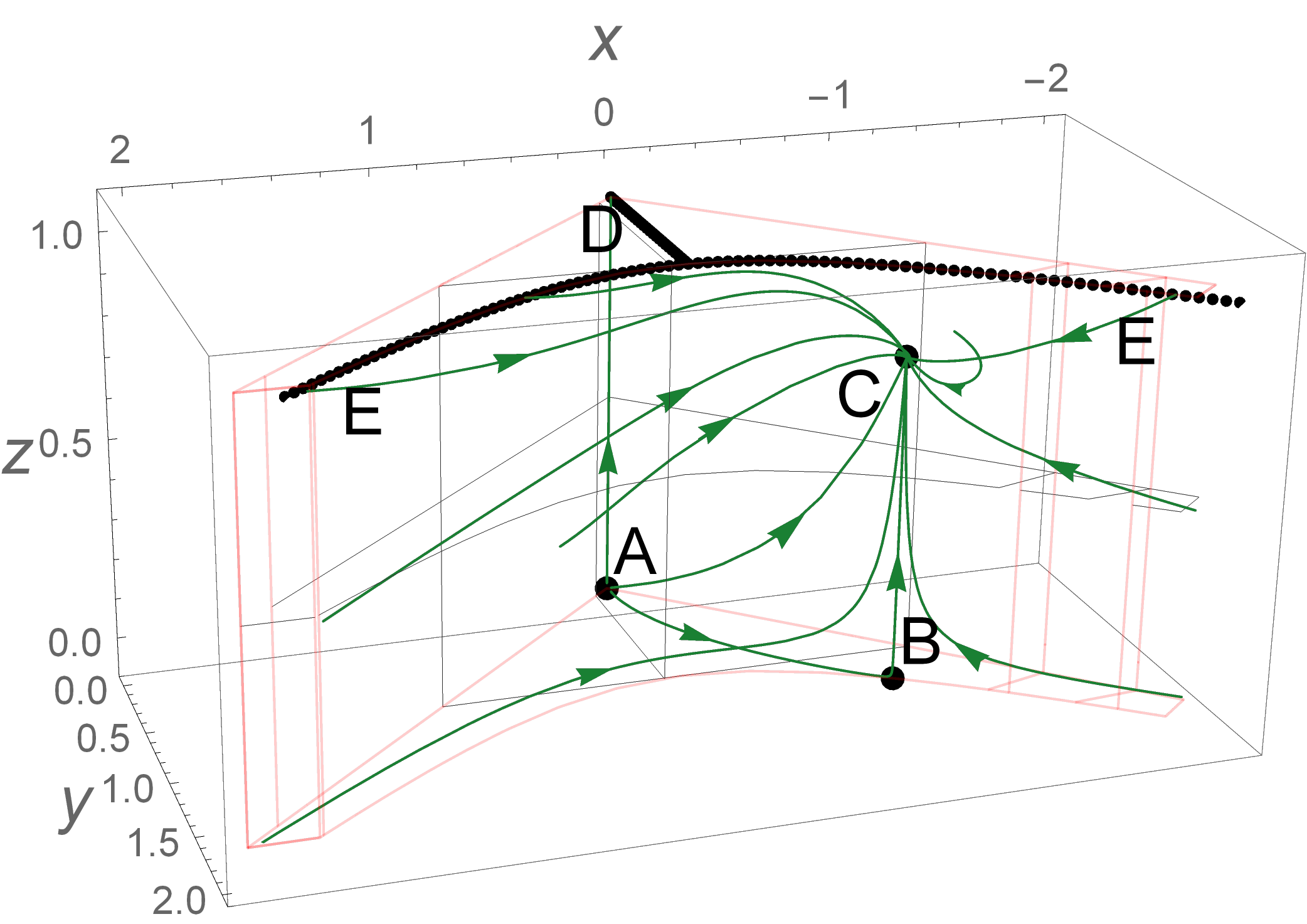}
    \caption{Phase portrait of the dynamical system (\ref{DS-1}) for $\alpha=-1$ and $\lambda=2$. For negative values of the parameter $\alpha$, the global attractor is the critical point $C$, representing a phantom accelerating solution for which $w_{\rm tot}<-1$.}
    \label{fig:PhS-neg-alpha}
\end{figure}

The critical point $C$ belongs to the phase space only if $\alpha<0$, i.e., only when there is an energy transfer from the phantom DE scalar field to the DM fluid.
This critical point represents a DE-dominated accelerating solution ($\Omega_\phi=1$, $w_{\rm tot}<-1$).
Because all eigenvalues are negative, point $C$, whenever exists, is a global attractor to which all orbits converge asymptotically (see Fig.~\ref{fig:PhS-neg-alpha}).

\begin{figure}
    \centering
    \includegraphics[width=0.5\textwidth]{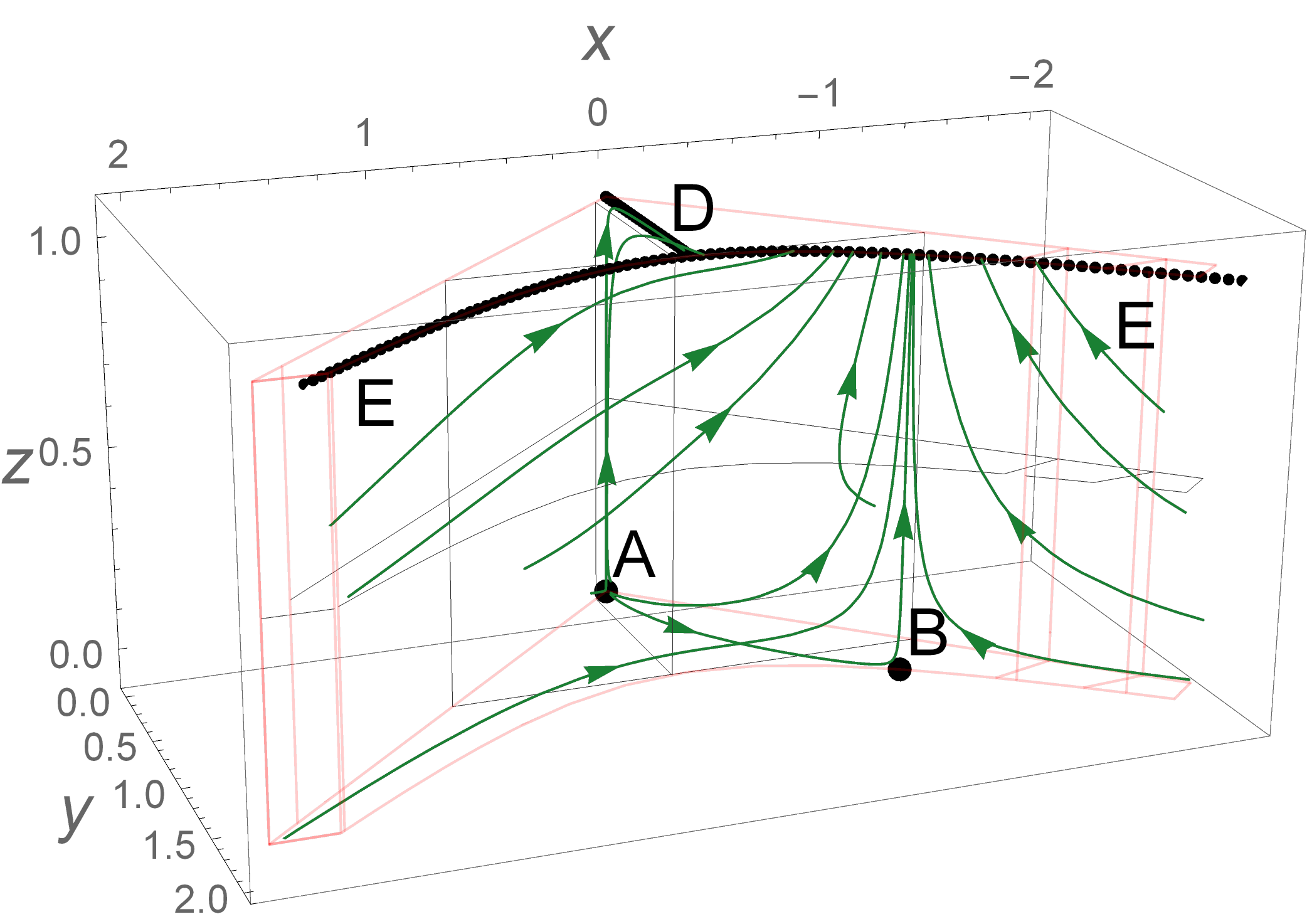}
    \caption{Phase portrait of the dynamical system (\ref{DS-1}) for $\alpha=1$ and $\lambda=2$. For positive values of the parameter $\alpha$, the attractor is the critical line $E$, representing a phantom accelerating solution for which $w_{\rm tot}<-1$.  }
    \label{fig:PhS-pos-alpha}
\end{figure}
\begin{figure*}
    \centering
    \includegraphics[width=0.32\textwidth]{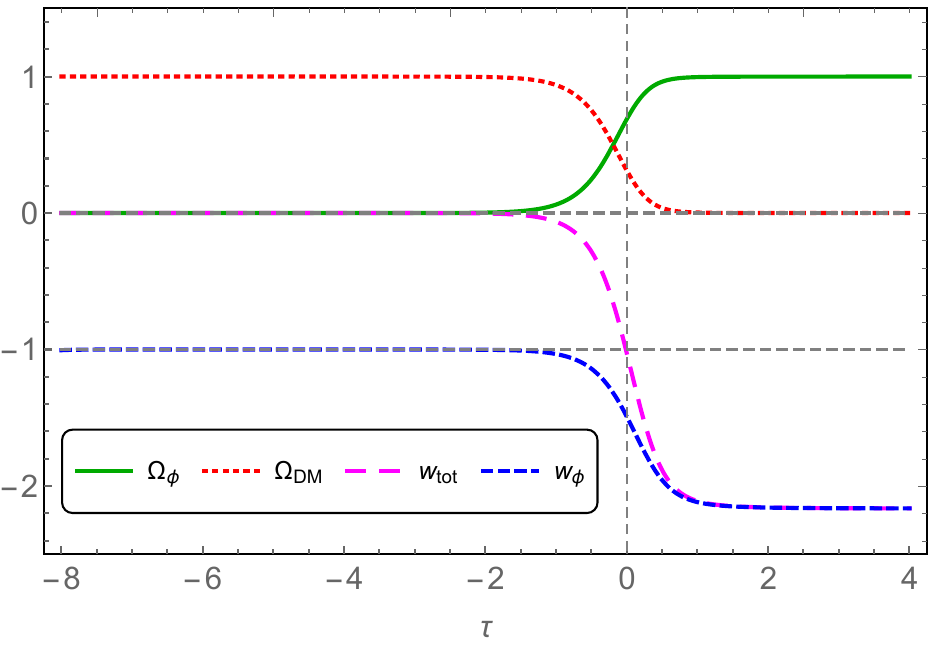}
    \includegraphics[width=0.32\textwidth]{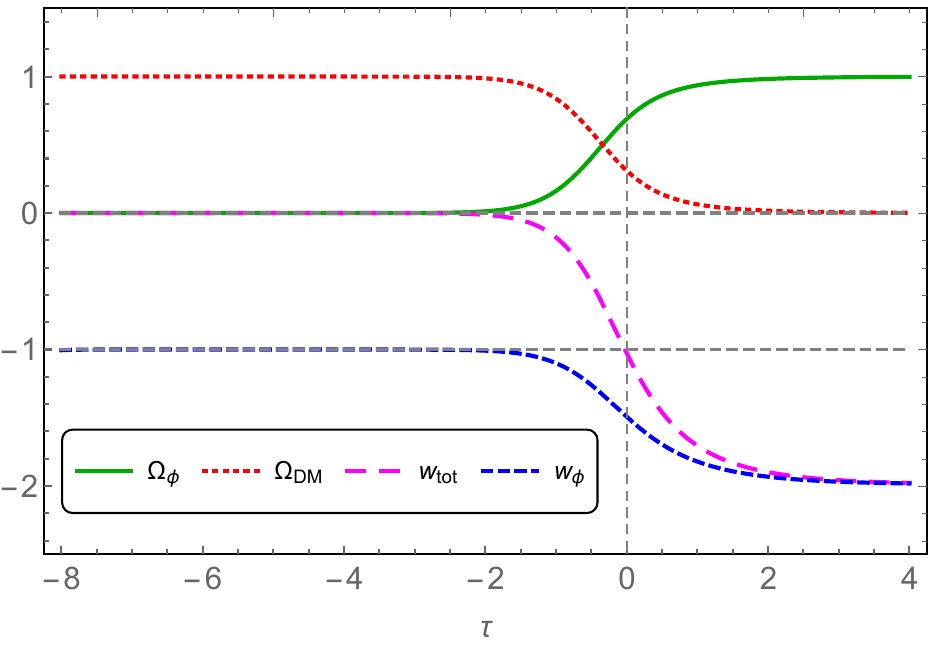}
    \includegraphics[width=0.32\textwidth]{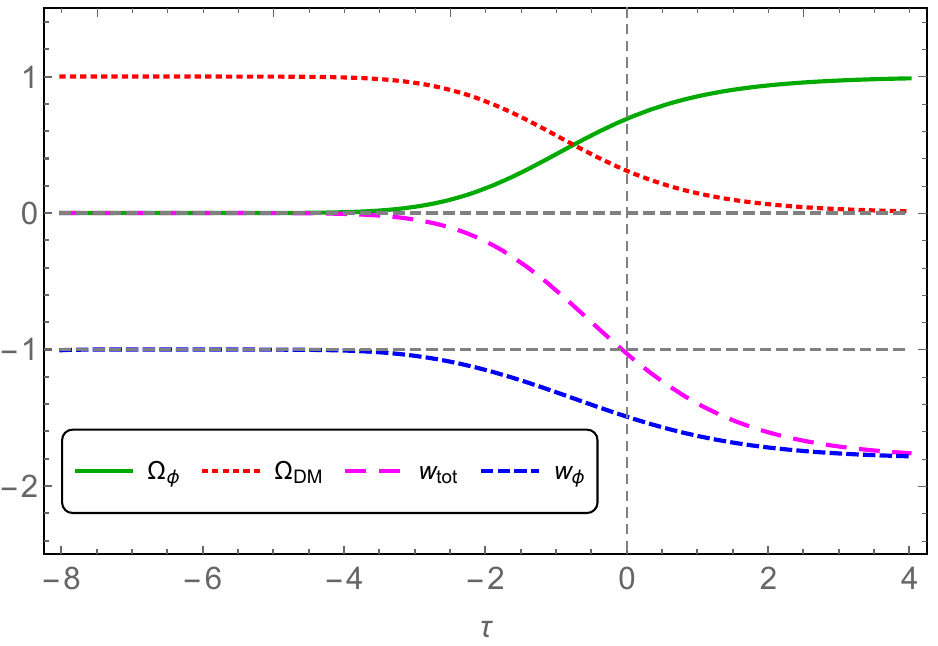}
    \caption{Evolution of $\Omega_\phi$, $\Omega_{\rm DM}$, $w_\phi$, and $w_{\rm tot}$ for $\alpha=1$ and different values of $\lambda=1.88$, $2.08$, and $2.95$ (left to right panels). Initial conditions at $\tau=-8$ (redshift of about $3000$) are such that all solutions yield $\Omega_\phi(0)\approx 0.69$ and $w_{\rm tot}(0)\approx-1.03$. In all cases, the matter-dominated era is long enough to allow for structure formation. Contrarily to the uncoupled scenario, the smaller the parameter $\lambda$, the more phantom the solution is.}
    \label{fig:alpha1-diff-lambda}
\end{figure*}
The critical line $D$, consisting of a continuous set of critical points, exists for all values of the parameters $\alpha$ and $\lambda$.
Critical points with $0 <y < 1$ correspond to scaling solutions for which the ratio $\Omega_{\rm DM}/\Omega_\phi$ is nonzero. 
Furthermore, these scaling solutions are accelerated if $y>1/\sqrt3$. 
However, as shown in the Appendix using the center manifold theory, these critical points are unstable for $0\leq y<1$ (and any value of $\alpha$), i.e., they cannot correspond to a final state for which $\Omega_{\rm DM}/\Omega_\phi \neq 0$ and, therefore, unfortunately, they cannot solve the cosmic coincidence problem. 
On the other hand, the critical point with $y=1$ is an attractor (for $\alpha>0$), but does not correspond anymore to a scaling solution, since for it the DM density parameter vanishes ($\Omega_{\rm DM} = 0$). 

Finally, the non-isolated critical points of the line $E$, which exist for all values of the parameters $\alpha$ and $\lambda$, correspond to DE-dominated accelerating solutions ($\Omega_\phi=1$, $w_{\rm tot}<-1$).
As shown in the Appendix, these critical points are stable for positive $\alpha$, i.e., for such values of $\alpha$, all orbits converge asymptotically to these points (see Fig.~\ref{fig:PhS-pos-alpha}).

Among all possible phase-space orbits, only a set reproduces the succession of cosmological eras observed in the late-time evolution of the universe, namely, an era dominated by matter, long enough to allow for structure formation, followed by an era of accelerated expansion.   

Let us focus our attention on these cosmologically relevant orbits.
They must pass close to the critical point $A$ to guarantee the existence of a long enough matter-dominated era and then proceed to the final state, which is the critical point $C$ if $\alpha<0$ or the critical line $E$ if $\alpha>0$ (see Figs.~\ref{fig:PhS-neg-alpha} and \ref{fig:PhS-pos-alpha}).

For $\alpha<0$, the final state $C$ has the same $x$ and $y$ coordinates as the global attractor $A_1$ of the uncoupled case (see section~\ref{Sec-2}), implying that the asymptotic values of the quantities $\Omega_\phi$, $\Omega_{\rm DM}$, $w_\phi$, and $w_{tot}$, which, recall, depend only on $x$ and $y$, coincide in both cases.
Therefore, the introduction of the interaction term (\ref{interaction}) between DE and DM does not seem to influence the late-time evolution of the universe if $\Gamma<0$ ($\alpha<0$).

Quite different is the situation for $\alpha>0$. Here, the $x$ and $y$ coordinates of the final state $E$ may not coincide with the corresponding coordinates of the global attractor of the uncoupled case, leading to different late-time behaviors of $\Omega_\phi$, $\Omega_{\rm DM}$, $w_\phi$, and $w_{tot}$.
More specifically, orbits that, after passing near $A$, proceed first to the vicinity of $B$ and only then head to the final state at the critical line $E$, correspond to cosmological solutions similar to those obtained in the uncoupled case.
On the contrary, orbits that, after passing near $A$, proceed directly to the final state, end up on a critical point not lying above $B$, and consequently yield different late-time behaviors for the above-referred physical quantities.
These two types of orbits can be easily identified in Fig.~\ref{fig:PhS-pos-alpha}.

Let us analyze in more detail the late-time behavior of $\Omega_\phi$, $\Omega_{\rm DM}$, $w_\phi$, and $w_{\rm tot}$ for $\alpha>0$, considering three examples for which $\alpha=1$ and $\lambda=1.88$, $2.08$, and $2.95$ (see Fig.~\ref{fig:alpha1-diff-lambda}).
In all cases, we choose initial conditions at $\tau=-8$ such that all solutions yield $\Omega_\phi(0)\approx 0.69$ and $w_{\rm tot}(0)\approx-1.03$, in agreement with observations\footnote{From Eq.~(\ref{tau}), and considering that $z\approx 0$ near the critical point $A$, it follows that a matter-dominated era starting at a redshift of about $3000$ corresponds to $\tau\gtrsim-8$. To guarantee that $\Omega_\phi(0)\approx 0.69$ and $w_{\rm tot}(0)\approx-1.03$, the initial conditions at $\tau=-8$ were chosen to be: $x_i=3.49\times10^{-7}$, $y_i=20x_i$, and $z_i=10^{-7}$ for $\lambda=1.88$; $x_i=7.61\times10^{-7}$, $y_i=20x_i$ and $z_i=10^{-5}$ for $\lambda=2.08$; $x_i=1.32\times10^{-5}$,  $y_i=20x_i$, and $z_i=10^{-3}$ for $\lambda=2.95$.}.

For $\lambda=1.88$, the corresponding phase-space orbit passes very close to $B$, implying that the $x$ coordinate at the final state is $x_f\approx-0.76$, almost the same value as the $x$ coordinate of the critical point $B$, $x_B\approx-0.77$. Therefore, the total equation-of-state parameter has similar asymptotic values in the coupled and uncoupled cases, namely, $w_{\rm tot}\approx-2.16$ and $-2.18$, respectively.

For $\lambda=2.08$, the orbit does not pass near $B$, heading directly to a final state with an $x$ coordinate, $x_f\approx-0.70$, quite different from the $x$ coordinate of the critical point $B$,  $x_B\approx-0.85$. This circumstance implies that the asymptotic value of the total equation-of-state parameter, $w_{\rm tot}\approx-1.98$, becomes noticeably higher than the corresponding value in the uncoupled case, $w_{\rm tot}\approx-2.44$. In other words, a direct energy transfer from DM to phantom DE results in a solution less phantom. 

This difference between the coupled and uncoupled cases is more pronounced in the case $\lambda=2.95$, in which the corresponding phase-space orbit heads even more directly to the final state, yielding $x_f\approx-0.62$ and $w_{\rm tot}\approx-1.77$, while in the uncoupled case these values are $x_B\approx-1.20$ and $w_{\rm tot}\approx-3.90$.

The above results can be summarized as follows. In the uncoupled phantom model, the equation-of-state parameter's asymptotic value is determined solely by $\lambda$, namely,  $w_{\rm tot}=-1-\lambda^2/3$ (see Table~\ref{first-table-unc-potI}). Thus, as $\lambda$ increases, $w_{\rm tot}$ becomes more negative, leading to a solution more phantom. When the direct coupling (\ref{interaction}) with $\Gamma>0$ ($\alpha>0$) is introduced in the evolution equations, the equation-of-state parameter depends not only on $\lambda$ but also on the energy exchange between DE and DM. As a result, the asymptotic behavior of $w_{\rm tot}$ is reversed when compared to the uncoupled case, namely, $w_{\rm tot}$ increases with increasing $\lambda$, making the solution less phantom.

To conclude this section, let us point out that outside the phase space $\overline{R}_3$, on the boundary of the region $R_3$, defined in Eq.~(\ref{R3}), for $\alpha<0$, there are two more critical points, namely, $F_{\pm}(\pm1,0,3/(3-2\alpha))$. Such critical points attract a set of orbits, which, instead of converging to the phantom final state $C$, end up in the unphysical region $y<\pm x$. Such a circumstance would require, in general, the imposition of an additional constraint, either on the phase space or on the parameter space. However, as we have checked numerically, none of the orbits of cosmological relevance (those that guarantee a long enough matter-dominated era by passing near the critical point $A$) is attracted to these unphysical critical points, so we can safely ignore them.

\section{Conclusions}
\label{Sec-conclusions}

Cosmological models with an energy exchange between DE and DM have gathered noteworthy attention in the scientific community because of their rich phenomenological consequences. It is essential to note that the dynamics of such coupled cosmological models depend significantly on the nature of the dark components and the energy exchange rate between them; since none of these features is currently known, there is a large freedom in constructing coupled quintessence and phantom cosmological models.

In this article, we have considered an interacting scenario between a pressureless DM fluid and a phantom DE scalar field with an exponential potential in which the interaction function is motivated by the warm inflationary paradigm. More specifically, we have assumed the interaction term $Q$ to be of the dissipative type, $Q=\Gamma \dot{\phi}^2$, where $\Gamma$ is a dissipation coefficient determined by local properties of the dark-sector interactions. In a first, simplified approach to this model, we have chosen $\Gamma$ to be constant, leaving more general cases for future work. 

To understand the salient features of the interacting dynamics, we have first reviewed the dynamics of the uncoupled phantom DE cosmological model. Using methods of qualitative analysis of dynamical systems, we have shown that this uncoupled model admits a set of cosmologically relevant solutions that reproduce the succession of cosmological eras observed in the late-time evolution of the universe, namely, an era dominated by matter, long enough to allow for structure formation, followed by the current era of accelerated expansion.
However, contrarily to the corresponding uncoupled quintessence model \cite{Copeland:1998}, there are no scaling attractor solutions; the unique late-time attractor in the uncoupled phantom model describes a Universe completely dominated by phantom DE.
Note also that the asymptotic value of the total equation-of-state parameter $w_{\rm tot}=-1-\lambda^2/3$ is fixed solely by the choice of the parameter $\lambda$, related to the steepness of the potential of the scalar field.

When the aforementioned direct coupling between the phantom DE scalar field and the DM fluid is introduced in the evolution equations, the dynamical system's phase space structure changes considerably from what we observe in the uncoupled scenario, allowing for a different late-time behavior of the solutions.

First, the dynamical system of the coupled model admits a set of non-isolated critical points corresponding to accelerated scaling solutions, i.e., solutions for which $\Omega_{\rm DM}/\Omega_\phi \neq 0$ and $w_{\rm tot}<-1/3$ (see Table~\ref{first-table-potI}). However, since these critical points are unstable for any value of $\alpha$, they do not correspond to a final state of the universe's evolution and hence the coincidence problem cannot be solved within this particular model.
The absence of scaling attractor solutions is quite common in coupled phantom cosmological models\footnote{It might be interesting to note that in the context of coupled DM-DE scenarios in which the coupling function has sign shifting nature and DE may behave like a phantom fluid (without being a phantom scalar field) can offer accelerated scaling solutions which are stable \cite{Halder:2024uao}, and hence, they can alleviate the coincidence problem.}. Indeed, such solutions were proven not to exist for various interaction terms between DM and phantom DE, mainly considering an exponential potential for the phantom scalar field \cite{Guo:2004xx,Chen:2008ft,Leon:2009dt,Shahalam:2017fqt}, but also power-law and hyperbolic potentials \cite{Leon:2009dt,Zonunmawia:2017ofc}. However, scaling attractor solutions have been found for specific interaction terms \cite{Guo:2004xx,Chen:2008ft}, although in one instance at the expense of fine-tuning the model parameters \cite{Chen:2008ft}. It is unclear why phantom DE appears to favor the absence of scaling attractor solutions, even in models where the interaction term allows for a substantial transfer of energy from the phantom DE to the DM fluid. This is an issue that warrants further investigation.

Second, for $\alpha>0$ (indicating an energy transfer from DM to phantom DE), the asymptotic value of the total equation-of-state parameter $w_{\rm tot}$ is no longer fixed uniquely by the choice of $\lambda$, as in the uncoupled scenario. Instead, it also depends on the energy exchange between the dark components. Our dynamical system analysis shows that, in the coupled scenario, the phase-space orbits asymptotically converge to a critical line, ending up at different points of this line. To each of these points corresponds a different asymptotic value of $w_{\rm tot}$, namely, $w_{\rm tot}=-1-2x_f^2$, where $x_f$ is the $x$ coordinate of the point. The asymptotic values of $w_{\rm tot}$ are such that higher values of $\lambda$ correspond to solutions less phantom, which is exactly the opposite of what happens in the uncoupled scenario. Therefore, a direct energy transfer from DM to DE through a dissipative term inspired by warm inflation significantly alters the late-time behavior of the phantom DE cosmological model. 

Based on the outcomes of the present work, we deem it important to further explore the coupled phantom DE cosmological model inspired by warm inflation. 
In particular, models with a dissipation coefficient $\Gamma$ depending on both the phantom scalar field and the dark-matter energy density, as well as with different potentials for the phantom scalar field, will be considered in future work. It will be interesting to examine whether these coupled phantom DE cosmological models can lead to stable accelerating scaling solutions, thus, solving the cosmic coincidence problem.

\begin{acknowledgments}
We thank the referee for some useful comments that improved the quality of the manuscript.
SH acknowledges the financial support from the University Grants Commission (UGC), Govt. of India (NTA Ref. No: 201610019097).
SP and TS acknowledge the financial support from the Department of Science and Technology (DST), Govt. of India under the Scheme   ``Fund for Improvement of S\&T Infrastructure (FIST)'' (File No. SR/FST/MS-I/2019/41).
PS acknowledges support from Funda\c{c}\~ao para a Ci\^encia e a Tecnologia (Portugal) through the research grants doi.org/10.54499/UIDB/04434/2020 and doi.org/10.54499/UIDP/04434/2020.
\end{acknowledgments}

\appendix*

\section{Stability of the critical lines $D$ and $E$ \label{Sec-appendix}}

In this appendix, we investigate the stability of the critical lines $D(0,y,1)$ and $E(x,\sqrt{1+x^2},1)$.

Let us start with the critical line $E(x,\sqrt{1+x^2},1)$, considering, to that end, a specific point $E(x_c,\sqrt{1+x_c^2},1)$, where $-\infty<x_c<+\infty$.

The Jacobian matrix of the dynamical system (\ref{DS-1}), given by
\begin{equation}
 J_E=\begin{pmatrix}
     -2 \alpha x_c^2  & 2 \alpha x_c \sqrt{1+x_c^2}  & 3 (1+x_c^2) (x_c+\frac{\sqrt6}{6} \lambda )
      \\
     0 & 0 & 3 x_c \sqrt{1+x_c^2} (x_c+\frac{\sqrt6}{6} \lambda) \\
     0 & 0 & -2 \alpha x_c^2
    \end{pmatrix},
   \label{jacA}
\end{equation}
has the eigenvalues
\begin{equation}
    \lambda_1=0, \quad \lambda_{2,3}=-2 \alpha x_c^2.
\end{equation}

Since the above Jacobian matrix has only a zero eigenvalue, the non-isolated critical point $E(x_c,\sqrt{1+x_c^2},1)$ is normally hyperbolic, meaning that stability can be assessed within the linear theory.
Because the nonzero eigenvalues are negative for $\alpha>0$, this critical point is an attractor for such values of the parameter $\alpha$. 

Let us now turn to the analysis of the stability of the critical line $D(0,y,1)$, considering, to that end, a specific point $D(0,y_c,1)$, where $0\leq y_c<1$.

The Jacobian matrix of the dynamical system (\ref{DS-1}), given by
\begin{equation}
 J_D=\begin{pmatrix}
     \alpha(y_c^2-1) & 0 & \frac{\sqrt6}{2}\lambda y_c^2
      \\
     0 & 0 & \frac32 y_c (y_c^2-1) \\
     0 & 0 & 0
    \end{pmatrix},
   \label{jacA}
\end{equation}
has the eigenvalues
\begin{equation}
    \lambda_1=\alpha(y_c^2-1), \quad \lambda_{2,3}=0,
\end{equation}
to which correspond the (generalized) eigenvectors
\begin{equation}
    v_1=
    \begin{pmatrix}
    1 \\ 0 \\ 0
    \end{pmatrix},
   \quad v_2=
   \begin{pmatrix}
    0 \\ 1 \\ 0
    \end{pmatrix},
    \quad v_3=
    \begin{pmatrix}
    \frac{\sqrt6 \lambda y_c^2}{2\alpha(1-y_c^2)} \\ 0 \\ 1
    \end{pmatrix}.
    \label{gen_eig_vec}
\end{equation}

Since the Jacobian matrix $J_D$ has two zero eigenvalues, the linear theory is not enough to assess the stability of the critical point  $D(0,y_c,1)$ and, consequently, one has to resort to alternative methods.
Here, we choose the center manifold theory \cite{Carr:1982,Guckenheimer:1983,Bogoyavlensky:1985}.

To shift the critical point $D(0,y_c,1)$ to the origin of the coordinate system, we introduce new variables
\begin{equation}
    u=x, \quad v=y-y_c, \quad w=z-1,
\end{equation}
for which the dynamical system (\ref{DS-1}) becomes
\begin{subequations}
\begin{align}
 u_\tau & = -\alpha (1-y_c^2)u + \frac{\sqrt6}{2} \lambda y_c^2 w + f_1(u,v,w),
\\
 v_\tau & = -\frac32 y_c (1-y_c^2)w + f_2(u,v,w),
\\
 w_\tau & = f_3(u,v,w),
\end{align}
\end{subequations}
where $f_i=\mathcal{O}(u^2,v^2,w^2,uv,uw,vw)$, $i=1,2,3$.

Another change of variables, namely,
\begin{equation}
 \begin{pmatrix}
  u\\v\\w
 \end{pmatrix}
  = S
 \begin{pmatrix}
 U\\V\\W
 \end{pmatrix},
\end{equation}
where $S$ is a matrix whose columns are the generalized eigenvectors (\ref{gen_eig_vec}),
brings the dynamical system to the form
\begin{subequations}
\begin{align}
 U_\tau & = -\alpha (1-y_c^2)U + F_1(U,V,W),
\\
 V_\tau & = -\frac32 y_c (1-y_c^2)W + F_2(U,V,W),
\\
 W_\tau & = F_3(U,V,W),
\end{align}
\end{subequations}
where $F_i=\mathcal{O}(U^2,V^2,W^2,UV,UW,VW)$, $i=1,2,3$.

Before proceeding to the determination of the center manifold $U=h(V, W)$ and the flow on it, let us point out that along the $U$ direction, the orbits approach the critical point for $\alpha>0$ and move away from it for $\alpha<0$.

Now, the center manifold is a solution to the partial differential equation
\begin{align}
 & \frac{\partial h}{\partial V} \left[
 -\frac32 y_c (1-y_c^2) W
 +F_2\Big(h(V,W),V,W\Big) \right] \nonumber
\\
 &\hspace{5mm}+
 \frac{\partial h}{\partial W} F_3\Big(h(V,W),V,W\Big) +\alpha (1-y_c^2) h(V,W)\nonumber
\\
 &\hspace{5mm} -F_1\Big(h(V,W),V,W\Big)=0,
\end{align}
where $h(V,W)$ is defined on some neighborhood of the critical point with $h(0,0)$ and $\nabla h(0,0)=0$.

Searching for a solution to the above equation of the form
\begin{equation}
 h(V,W)=\sum_{j=2}^m \sum_{i=0}^j a_{i,j-i}V^i W^{j-i},
\end{equation}
where $a_{ij}$ are constants and $m\geq2$, we obtain, at lowest order in powers of $V$ and $W$,
\begin{equation}
 U=\frac{\sqrt6\lambda y_c}{\alpha(1-y_c^2)^2} VW
+ \frac{\sqrt6 \lambda y_c^2[3-\alpha(1-y_c^2)]}{2\alpha^2(1-y_c^2)^2} W^2.
\end{equation}

As both terms of the above solution vanish for $y_c=0$, for such a value of $y_c$ we have to extend our calculation to the third order in powers of V and W, obtaining
\begin{equation}
U=\frac{\sqrt6\lambda}{2\alpha}V^2W.
\end{equation}

The flow on the center manifold is determined by the differential equations
\begin{subequations}
\label{flow-V,W}
\begin{align}
V_\tau & =  -\frac32 y_c (1-y_c^2 )W,
\label{flow-V}
\\
 W_\tau & = \frac32 (1+y_c^2)W^2.
 \label{flow-W}
\end{align}
\end{subequations}

Again, the right-hand side of Eq.~(\ref{flow-V}) vanishes for $y_c=0$, requiring, for such a value of $y_c$, to extend the calculation to higher orders in powers of $V$ and $W$.
The flow on the center manifold is then, for $y_c=0$, determined by the differential equations
\begin{subequations}
\label{flow-V,W_zero}
\begin{align}
V_\tau & =  -\frac32 V W,
\label{flow-V_zero}
\\
 W_\tau & = \frac32 W^2.
 \label{flow-W_zero}
\end{align}
\end{subequations}

For $0<y_c<1$, taking into account that $W<0$ in the neighborhood of the critical point, it follows that both $V_\tau$ and $W_\tau$ are positive and, consequently, the orbits approach the critical point along the $W$-direction and drift in the direction of increasing $V$.
For the case $y_c=0$, taking also into account that $V>0$, we arrive at the very same conclusion.
Note that this result does not depend on the parameter $\alpha$, contrarily to the result obtained for the flow along the $U$-direction, which, as mentioned above, approaches the critical point for $\alpha>0$ and moves away from it for $\alpha<0$.

In terms of the original variables $x$, $y$, and $z$, the above results mean that, for $\alpha>0$, the orbits, when approaching the critical line $D(0,y,1)$, drift in the $y$-direction, towards the critical point $D(0,1,1)$, which coincides with the point $E(0,1,1)$ of the critical line $E(x,\sqrt{1+x^2},1)$.

This result, together with the stability analysis of the critical line $E(x,\sqrt{1+x^2},1)$, leads to the conclusion that for positive values of the parameter $\alpha$ the critical line $E(x,\sqrt{1+x^2},1)$ is an attractor, i.e., all orbits (except the heteroclinic ones, connecting critical points along the boundaries of the phase space) asymptotically converge to one of the points of this critical line.

\bibliography{biblio}

\end{document}